\documentstyle[multicol,prb,aps,psfig]{revtex}

\input epsf

\def\prb{Phys. Rev. B }
\def\prl{Phys. Rev. Lett. }

\def\be{\begin{equation}}
\def\ee{\end{equation}}
\def\ba{\begin{eqnarray}}
\def\ea{\end{eqnarray}}

\def\LSCO{La$_{2-x}$Sr$_x$CuO$_4$ }
\def\LSNO{La$_{2-x}$Sr$_x$NiO$_4$ }
\def\YBCO{YBa$_2$Cu$_3$O$_{7-\delta}$ }
\def\124{YBa$_2$Cu$_4$O$_8$ }

\def\BSCCO{Bi$_2$Sr$_2$CaCu$_2$O$_{8+\delta}$ }
\def\C60{A$_x$C$_{60}$ }
\def\LNSCOsc{La$_{1.6-x}$Nd$_{0.4}$Sr$_{x}$CuO$_{4}$ }
\def\LNSCOin{La$_{1.4-x}$Nd$_{0.6}$Sr$_{x}$CuO$_{4}$ }

\def\LNO{La$_{2}$Ni$_{2}$O$_{4.125}$ }

\def\LBSCO{La$_{1.875}$Ba$_{0.125-x}$Sr$_x$CuO$_4$ }

\def\ie{ {\it i.e.} }

\begin{document}

%\twocolumn[\hsize\textwidth\columnwidth\hsize\csname@twocolumnfalse\endcsname

\title
{Interacting Electrons on a Fluctuating String}

\author{Dror~Orgad}

\address
{Racah Institute of Physics, The Hebrew University, Jerusalem 91904, Israel}

\date{\today}
\maketitle 

\begin{abstract}

We consider the problem of interacting electrons constrained to move 
on a fluctuating one-dimensional string. An effective low-energy theory 
for the electrons is derived by integrating out the string degrees of 
freedom to lowest order in the inverse of the string tension and mass density, 
which are assumed to be large. We obtain expressions 
for the tunneling density of states, the spectral function and the optical 
conductivity of the system. Possible connections with the phenomenology of 
the cuprate high temperature superconductors are discussed.

\end{abstract}

\begin{multicols}{2}

\section{Introduction}
\label{intro}

The problem of interacting electrons moving in one dimension has attracted 
considerable attention since the early theoretical formulations of 
Tomonaga \cite{tomonaga50} and Luttinger \cite{luttinger63}.
The motivation for the continuous activity in this field has been twofold. 
First, this system offers a concrete realization of various non-Fermi 
liquid phenomena and is amenable to controlled theoretical treatments. 
As such it constitutes a unique theoretical laboratory for studying strong 
correlations. Secondly, in addition to the existence of various 
quasi-one-dimensional compounds such as organic Bachgaard salts 
\cite{bachgaard} and inorganic purple bronzes \cite{bronzes}, 
there is a growing set of higher dimensional systems whose experimentally 
observed behavior appears to be, nevertheless, quasi-one-dimensional. 
Among them one finds the high temperature superconductors 
\cite{pnas,zaanenscience,dimxover,frac,ando1d}, 
the manganites \cite{manganites} and quantum Hall systems 
\cite{lilly99,du99,foglerreview,qhsmectic1}. It is possible that 
the electronic structure of these materials is actually 
quasi-one-dimensional on a local scale.

The overwhelming majority of the studies of one-dimensional systems 
that have been carried out so far have assumed that the electrons 
move along static straight chains. A notable exception is the 
study of the smectic phase in quantum Hall samples 
\cite{qhsmectic1,fertigQH,mdfisherQH,halperin,qhsmectic2}. 
This theory presupposes 
the existence of a unidirectional charge density wave in the sample and 
focuses on the quantum mechanical description of the fluctuating chiral edge 
channels that develop on the boundaries between regions of different filling 
fraction. Each channel is characterized by a displacement field $Y$, 
representing the transverse displacement of the edge from its classical 
ground state position (taken to be along the $x$ axis), and a Luttinger 
field $\phi$, describing the phase of the electronic charge density modulation 
along the edge. In this problem the two are coupled through the term 
$\vec{J}\cdot\vec{A}$, which in the gauge $\vec{A}=By\hat{x}$, 
and using the fact that the edge current is $e\partial_t\phi$ 
translates into $(eB)Y\partial_t\phi$. As a result $Y$ and $\phi$ are 
canonically conjugated and thus do not constitute independent degrees of 
freedom. This is a fundamental difference between the quantum Hall edge and 
the problem of spinful interacting electrons on a fluctuating string which 
this paper explores.   

While the model of interacting electrons on a dynamical one-dimensional 
geometry possesses an independent theoretical appeal, we are particularly 
motivated by its possible relevancy to the physics of doped Mott insulators 
\cite{ourreview}. There is evidence that over a wide range of the phase 
diagram of these systems the doped holes spontaneously segregate into 
one-dimensional charged stripes which also form domain walls across which 
the background spin texture undergoes a $\pi$-phase shift. 

In some instances the observed stripe order is static, at least on the 
time scale of the experimental probe.
Elastic neutron diffraction measurements revealed static spin and charge 
stripe order in the quasi-two-dimensional insulating nickelate  \LNO 
\cite{LNiOtranq} and in the isostructural system \LSNO 
\cite{LNiOtranq,LSNOlee}. Similar results were obtained \cite{tranqLNSCO} 
for the non-superconducting relative \LNSCOin of the high-temperature 
superconductors. The charge order has since been confirmed by 
high-energy x-ray diffraction \cite{LNSCOxray}.
%Static planar domain walls 
%were also imaged using transmission electron microscopy in the 
%three-dimensional manganite \LCMO with $x=0.5$ \cite{manganites}.

Systems that exhibit static stripe order are typically insulators or bad 
conductors and stripe fluctuations seem to play an important role in turning 
them into superconductors. In the superconducting material \LNSCOsc the 
magnetic ordering temperature, $T_m$, and the superconducting transition 
temperature, $T_c$, are anti-correlated \cite{tranqLNSCOsc}.
For example, around $x=1/8$ where stripes are particularly stable, 
$T_m$ reaches a maximum while $T_c$ 
exhibits a substantial dip. The same effect has been observed in 
\LBSCO \cite{lbsco}. Static magnetic stripe signatures exist in 
the insulating spin-glass phase of the Nd-free \LSCO with 
$0.02\leq x\leq 0.05$ \cite{diagstripes} and also in the underdoped 
superconducting region $0.05<x\leq 0.12$ 
\cite{stripesLSCOsc1,stripesLSCOsc2}. However, near optimal doping and 
in the overdoped regime ($0.12< x\leq 0.25$), where quantum fluctuations 
are presumably larger, static ordering is averted and only dynamical stripe 
correlations have been found by inelastic neutron scattering experiments 
\cite{aeppliLSCOopt,yamadastripesLSCO}. Evidence for slowly 
fluctuating spin and charge order has also been detected in underdoped 
superconducting \YBCO \cite{mook1,mook2,araiYBCO,mook3}. 

Qualitative arguments and several estimates of the impact of stripes 
fluctuations on the inter-stripe couplings and the resulting phase 
diagram of the many-stripe system have been presented in 
Ref. \ref{kferef}. 
%In particular it was pointed out that such fluctuations 
%tend to degrade the coupling between charge density waves on neighboring 
%stripes while enhancing the Josephson coupling between them, thus 
%making superconductivity a more likely outcome. 
While the present paper does not deal with questions concerning the 
physics that emerges from such couplings 
%inter-stripe couplings and their 
%influence on the phase diagram of the fluctuating many-stripe system 
its purpose is to establish a quantum description of the single fluctuating 
one-dimensional electron gas as a basis for tackling these issues at a 
later stage.

In Sec. \ref{model} we introduce a model of interacting electrons
which are constrained to move on a fluctuating elastic string.  
We show how this constraint leads to a change in the metric and 
to the coupling of the electronic dynamics to effective scalar and 
vector potentials, which are determined by the string fluctuations
\cite{eduardo}. The model is then quantized and bosonized in 
Sec. \ref{quantization}. 

The electronic correlation functions of the resulting low-energy theory are
calculated in Sec. \ref{correlations} by integrating out
the string degrees of freedom under the assumption that it is stiff
and massive. This assumption enables us to generate a perturbative 
expansion in the inverse string tension and mass density.
%the average wavevector of string fluctuations (measured in terms 
%of the short distance cut-off associated with the string). 

As long as one is concerned with
correlations of order parameters which are projected on the
line defined by the string equilibrium configuration the
results are of the usual Luttinger type. 
The effects of fluctuations enter through the 
renormalization of the exponents of the power-laws that describe  
such correlations. If the number of particles is held fixed 
this renormalization is mainly due to the increase of the average
length of the string by its fluctuations. Consequently the relative
strength of the interactions between particles is increased compared
to their kinetic energy. If on the other hand a constant particle
density along the string is maintained this effect is absent and the 
renormalization is dominated by the attractive interaction induced  
by the exchange of elastic string waves.

More generally, however, the correlation functions depend  
on the position in the embedding plane, or equivalently, in Fourier 
space, on both the wave-vector component along the string axis, $k_x$,
and its component in the perpendicular direction $k_y$. An example 
is the single-hole spectral function. Under certain conditions we find
that it retains its Luttinger liquid structure in terms of $k_x$ and
the frequency $\omega$ but with interaction parameters which are
increasing functions of the momentum component $k_y$ along the
one-dimensional Fermi surface. This effect also leads to a weak logarithmic 
suppression of the tunneling density of states in addition to the 
renormalization of its power-law exponent as discussed above.

When the string fluctuates electrical current may flow in the $y$ direction
either because electrons are being dragged by the string or because charge
flows along segments of the string which are at an angle to the $x$ 
axis. The first process gives rise to a Drude-peak in the $y$ component
of the optical conductivity whose weight is smaller relative to the 
corresponding peak in $\sigma_{xx}$, by the ratio of the 
electronic and string mass densities . 
The second process results in a contribution which is linear 
in $\omega$ and whose oscillator strength is further reduced 
%by the ratio of the string waves velocity and the charge velocity 
in comparison to the Drude peak. However, this contribution survives even 
if the string is pinned and can not execute rigid translations, while the 
peak does not.

We end Section \ref{correlations} with a brief consideration of the
effects of placing the string inside a harmonic confining potential. 
As expected the potential predominantly sets a cross-over scale 
below which correlations are the same as of a regular Luttinger liquid. 

Finally, in Sec. \ref{discussion}, we discuss few
possible applications of our findings to measurements of the
cuprate high temperature superconductors.

\section{The Model}
\label{model}

We consider electrons that are constrained to move on a stretchable 
one-dimensional fluctuating string, embedded in a two-dimensional
plane. We assume that the projection of the string on the $x$ 
axis is of fixed length, $L_x$, and that it obeys periodic boundary 
conditions along this direction. We consider the limit of a stiff and 
massive string whose characteristic quantum and thermal 
fluctuations are smooth on the length-scale of
any short distance cutoff associated with its dynamics, for example 
the lattice spacing, $a$, in the case of stripes in doped Mott insulators. 
%Consequently we will work in the continuum limit and consider only 
%configurations that are smooth at the scale of this short distance cutoff. 
This assumption allows us to work in the continuum limit and to 
ignore overhangs. Consequently we describe the 
string using a function, $Y(x,t)$, with $0\leq x \leq L_x$, 
which represents its transverse oscillations relative to the string classical 
equilibrium configuration, which we take to be along the $x$ axis.

The string is characterized by two parameters: the linear mass 
density, $\rho$, and the tension, $\sigma$. In the case where the 
string models a stripe, $\rho$ is determined by the kinetic energy 
per unit length of the domain wall while $\sigma$ is given by the 
energy to create a unit length of the domain wall inside the Mott insulator. 
The string Lagrangian is therefore
\ba
\label{L_s}
\nonumber
L_S&=&\int_0^{L_x} dx \left[ {{\rho}\over{2}}
\left({\partial Y}\over{\partial t}\right)^2
-\sigma\sqrt{1+\left({\partial Y}\over{\partial x}\right)^2} \right ] \\
&\approx& \int_0^{L_x} dx \, {{\sigma}\over{2}}\left[\frac{1}{u^2}
\left({\partial Y}\over{\partial t}\right)^2 - 
\left({\partial Y}\over{\partial x}\right)^2 \right ] \; ,
\ea
where 
\ba
\label{u}
u=\sqrt{\frac{\sigma}{\rho}} \; ,
\ea
is the sound velocity on the string. The linearized approximation of the 
action is valid as long as the short distance cutoff for the string  
fluctuations \cite{acomment}, 
which we denote by, $a$, is large compared to the length 
scale $(\sigma\rho)^{-1/4}$ set by its parameters (here and throughout 
the paper we take $\hbar=1$.) Assuming that this condition holds 
we will use it below to carry out a perturbative calculation of the effective 
electronic action in the small dimensionless parameter
\be
\label{epsilon}
\epsilon=\left\langle\left(\frac{\partial Y}{\partial x}\right)^2\right
\rangle=\frac{1}{2\pi a^2\sqrt{\sigma\rho}} \; .
\ee
This can be interpreted as the average (squared) slope of the string relative 
to its equilibrium configuration or as the average fluctuations-induced local 
dilatation of the string, since $(\partial Y/\partial x)^2=(ds)^2/(dx)^2-1$, 
where $ds$ is the length element along the string.

Next we consider the $N_e$ electrons that move on the string.
The position of the $i$-th electron is given by its Cartesian 
coordinates in the plane 
\ba
\label{r_i}
\vec{r_i}(t)=[x_i(t),y_i(t)]=[x_i(t),Y(x_i(t),t)] \; ,
\ea
where the last equality expresses the constraint that confines the
particles to the string. Its velocity is given by
\be
\label{v_i}
\vec{v_i}=\frac{d\vec{r_i}}{dt}=\left[\frac{dx_i}{dt}\, , \, 
\left.\frac{\partial Y}{\partial x}\right|_{x=x_i}
\frac{dx_i}{dt}+\left.\frac{\partial Y}{\partial t}\right|_{x=x_i}\right] \; ,
\ee
which demonstrates that the electrons change their position in the plane 
either by moving along the string or by being dragged by it in the 
$y$-direction.

Including the coupling of the particles to 
an external electromagnetic field, with scalar potential $A_0$ and vector 
potential $\vec{A}$, the electronic Lagrangian is given by
\ba
\label{L_e1}
\nonumber
L_e=&&\sum_{i=1}^{N_e}\left[ \frac{1}{2}m\vec{v_i}^2 + eA_0(\vec{r_i},t) 
-\frac{e}{c}\vec{v_i}\cdot\vec{A}(\vec{r_i},t)\right] \\
&&-\frac{1}{2}\sum_{i\neq j=1}^{N_e}V(|\vec{r_i}-\vec{r_j}|) \; ,
\ea
where $m$ and $-e<0$ are the electronic mass and charge respectively, and 
$V$ is the pair interaction, which in the following we take to be short 
ranged. Using relations (\ref{r_i}) and (\ref{v_i}) this Lagrangian 
can also be written as 
\ba
\label{L_e2}
\nonumber
L_e=&&\frac{m}{2}\sum_{i=1}^{N_e} g(x_i,t)\left(\frac{dx_i}{dt}\right)^2 \\
\nonumber
&& +\sum_{i=1}^{N_e} \left[ e {\cal A}_0 (x_i,t) 
-\frac{e}{c}\frac{dx_i}{dt}{\cal A}_1 (x_i,t)\right] \\
&&-\frac{1}{2}\sum_{i\neq j=1}^{N_e}V(x_i,x_j) \; , 
\ea
where 
\be
\label{g}
g(x,t)=1+\left(\frac{\partial Y}{\partial x}\right)^2 \; ,
\ee
is the metric induced by the string in the coordinate system of the $x_i$s. 
The string fluctuations also participate in determining the effective gauge 
potentials for the $x_i$ degrees of freedom 
\ba
\label{Aeff}
\nonumber
{\cal A}_0(x,t)=&&\frac{m}{2e}\left(\frac{\partial Y}
{\partial t}\right)^2 \\
\nonumber
&&+A_0[x,Y(x,t),t] - \frac{1}{c}\frac{\partial Y}{\partial t}A_y[x,Y(x,t),t] 
\;, \\
\nonumber 
{\cal A}_1(x,t)=&&-\frac{mc}{e}\frac{\partial Y}{\partial x}
\frac{\partial Y}
{\partial t} \\
&&+A_x[x,Y(x,t),t] + \frac{\partial Y}{\partial x}A_y[x,Y(x,t),t] \; .
\ea
Finally the pair interaction is 
\be
\label{Vpair}
V(x,x')=V\left(\sqrt{(x-x')^2+[Y(x,t)-Y(x',t)]^2}\right) \; .
\ee

Equations (\ref{L_e2})-(\ref{Vpair}) establish a description of the electronic 
physics in Cartesian coordinates. Although simple in appearance this 
formulation poses difficulties when one attempts to quantize the model. 
The source of the problem is the space-time dependence of the metric 
that appears in the kinetic term of the Lagrangian. Upon quantization
it leads to operator ordering ambiguities. In order to circumvent this 
problem we parameterize the positions of the particles on the string 
using the arc-length variable  
\be
\label{arc}
\ell(x,t)=\int_0^x dx' \sqrt{g(x',t)} \; .
\ee
Denoting by $\ell_i=\ell(x_i,t)$ the position of the $i$-th particle 
along the string we find
\be
\label{xtol}
\frac{dx_i}{dt}=\frac{1}{\sqrt{g(x_i,t)}}\left[ \frac{d\ell_i}{dt}- 
\frac{\partial\ell}{\partial t}(x_i,t)\right] \; ,
\ee
where
\be
\label{partiall}
\frac{\partial\ell}{\partial t}(x,t)=\int_0^x dx' \frac{1}{2\sqrt{g(x',t)}}
\frac{\partial g(x',t)}{\partial t} \; .
\ee
In terms of the new coordinates the electronic Lagrangian appears as
\ba
\label{newL_e}
\nonumber
L_e=&&\sum_{i=1}^{N_e}\left[ \frac{m}{2} \left(\frac{d\ell_i}{dt}\right)^2 
 + e \tilde{\cal A}_0(\ell_i,t) 
-\frac{e}{c}\frac{d\ell_i}{dt}\tilde{\cal A}_1(\ell_i,t)\right] \\
&&-\frac{1}{2}\sum_{i\neq j=1}^{N_e} V[x(\ell_i),x(\ell_j)] \; ,
\ea 
and the gauge potentials are 
\ba
\label{newA}
\nonumber
\tilde{\cal A}_0(\ell,t)=&&{\cal A}_0[x(\ell),t]+
\frac{m}{2e}\left(\frac{\partial\ell}{\partial t} [x(\ell),t]\right)^2 \\
\nonumber
&&+\frac{1}{c}\frac{1}{\sqrt{g[x(\ell),t]}}
\frac{\partial\ell}{\partial t}[x(\ell),t]{\cal A}_1[x(\ell),t] \; , \\
\tilde{\cal A}_1(\ell,t)=&&\frac{mc}{e}\frac{\partial\ell}{\partial t} 
[x(\ell),t]+\frac{1}{\sqrt{g[x(\ell),t]}}{\cal A}_1[x(\ell),t] \; .
\ea

In this form the problem reduces to a system of particles in flat 
1+1 space-time (whose spatial extent is time dependent) 
which interact with each other via the pair potential $V$ 
and couple to the gauge potentials $\tilde{\cal A}_{\mu}$, which are 
functions of the string configuration $Y[x(\ell),t]$. The price we pay 
for the simple form of the kinetic term is the fact that these gauge 
potentials are non-local since they are functions of $\partial\ell/\partial t$ 
defined in Eq. (\ref{partiall}).

As a last step towards the quantization of the problem we identify the 
electronic Hamiltonian in the arc-length coordinate. The conjugate momentum 
to $\ell_i$ is $p_i=\partial L_e/\partial(d\ell_i/dt)=m(d\ell_i/dt)
-(e/c)\tilde{\cal A}_1(\ell_i,t)$ and the classical electronic Hamiltonian is
\ba
\label{H_e}
\nonumber 
H_e=&&\sum_{i=1}^{N_e}\left[\frac{1}{2m}\left(p_i+\frac{e}{c}\tilde{\cal A}_1
(\ell_i,t)\right)^2-e\tilde{\cal A}_0(\ell_i,t)\right] \\
&&+\frac{1}{2}\sum_{i\neq j=1}^{N_e} V[x(\ell_i),x(\ell_j)] \; .
\ea

\section{Quantization and Bosonization}
\label{quantization}

Quantizing the model defined by Eq. (\ref{H_e}) is a straightforward task, 
analogous to the quantization of a conventional electron gas. 
The resulting second quantized Hamiltonian for the many body-system, 
written in terms of the electronic fields $\tilde\psi_{\sigma}(\ell,t)$ 
for the two possible spin polarizations, is 
\ba
\label{H_eQ}
\nonumber
H_e&&=\!\int_0^{L(t)}\!\! d\ell\sum_{\sigma=\pm}\!\left[
%\tilde\psi_{\sigma}^{\dagger}i\partial_t\tilde\psi_{\sigma} \\
%\nonumber
%&&-
\tilde\psi_{\sigma}^{\dagger}\frac{1}{2m}\left(\!-i\partial_{\ell}
+\!\frac{e}{c}\tilde{\cal A}_1\right)^2\!\tilde\psi_{\sigma}
-e\tilde{\cal A}_0\tilde\psi_{\sigma}^{\dagger}\tilde\psi_{\sigma}\right] \\
\nonumber
&&+\frac{1}{2}\int_0^{L(t)}\!\!d\ell d\ell'\!\sum_{\sigma,\sigma'=\pm}\!
V_{\ell,\ell'}^{\sigma,\sigma'}\tilde\psi_{\sigma}^{\dagger}(\ell)
\tilde\psi_{\sigma'}^{\dagger}(\ell')\tilde\psi_{\sigma'}(\ell')
\tilde\psi_{\sigma}(\ell) \, , \\
\ea
where $L(t)=\int_0^{L_x}dx\sqrt{g(x,t)}$ is the instantaneous length of the 
string and where the spatial part of $V_{\ell,\ell'}^{\sigma,\sigma'}
=V_{\sigma,\sigma'}V[x(\ell),x(\ell')]$ takes the form Eq. (\ref{Vpair}). 
At this stage we are still considering the string degrees of freedom 
as a classical field.

%The gauge potentials ${\cal A}_0$ and ${\cal A}_1$ are now operators 
%given by the appropriately symmetrized versions of Eqs. (\ref{Aeff}) 
%and (\ref{newA}). This symmetrization is needed in order to ensure 
%that they are Hermitian, since after the quantization of the string dynamics 
%$\frac{\partial Y}{\partial t}$ and $Y$ no longer commute.

Despite its apparent simplicity the formulation of the problem in the 
arc-length parameterization, Eq. (\ref{H_eQ}), suffers from two significant 
shortcomings. The first is the non-local nature of the gauge
potentials $\tilde{\cal A}_\mu$. The second problem stems from the fact
that experiments measure the electronic correlation functions as functions 
of the coordinates in the embedding plane and not of the arc-length 
along the string. It is therefore desirable to rewrite the theory 
in terms of the projected coordinate $x$. It turns out that by doing
so one also remedy the first difficulty.  

To this end we note that the spatial and temporal derivatives transform as 
\ba
\label{transd}
\left.\frac{\partial}{\partial\ell}\right|_t&=&
\frac{1}{\sqrt{g}}\left.\frac{\partial}
{\partial x}\right|_t \; , \\
\left.\frac{\partial}{\partial t}\right|_l&=&
\left.\frac{\partial}{\partial t}\right|_x 
-\frac{1}{\sqrt{g}}\left.\frac{\partial \ell}{\partial t}\right|_x
\left.\frac{\partial}{\partial x}\right|_t \; ,
\ea
and the integration with its measure according to 
\ba
\label{transint}
\int_0^{L(t)} d\ell \rightarrow \int_0^{L_x} dx \sqrt{g} \; .
\ea
Since the fermionic fields in the arc-length parameterization obey the 
anti-commutation relations
\ba
\label{acl}
\nonumber
\left\{\tilde\psi_{\sigma}(\ell,t),\tilde\psi_{\sigma'}^{\dagger}
(\ell',t)\right\}&=&\delta(\ell-\ell')\delta_{\sigma,\sigma'} \\
&=&\frac{1}{\sqrt{g(x,t)}}\delta(x-x')\delta_{\sigma,\sigma'} \; , 
\ea
we rescale the transformed fermionic fields according to
\be
\label{transpsi}
\psi_{\sigma}(x,t)=[g(x,t)]^{1/4}\tilde\psi_{\sigma}[\ell(x,t),t] \; ,
\ee
such that the new fields obey the canonical anti-commutation relations. 
In this way we also have $\rho_{\sigma}(x,t)=\psi_{\sigma}^
{\dagger}(x,t)\psi_{\sigma}(x,t)$ as the projected electronic 
densities in the length element $dx$. 

The result for the transformed electronic Lagrangian is then given by  

\end{multicols}
\widetext

\noindent
\setlength{\unitlength}{1in}
\begin{picture}(3.375,0)
  \put(0,0){\line(1,0){3.375}}
  \put(3.375,0){\line(0,1){0.08}}
\end{picture}
\ba
\label{Ltrans}
\nonumber
L_e&=&\int_0^{L_x} \! dx \sum_\sigma\left[
i\psi_\sigma^{\dagger}\partial_t\psi_\sigma
-\frac{1}{2m}\psi_\sigma^{\dagger}g^{-\frac{1}{4}}\left(-i\partial_x+
\frac{e}{c}{\cal A}_1\right)g^{-\frac{1}{2}}\!\left(-i\partial_x+
\frac{e}{c}{\cal A}_1\right)g^{-\frac{1}{4}}\psi_\sigma+e{\cal A}_0\rho_\sigma
-\frac{1}{2}g^{-\frac{1}{2}}\sum_{\sigma'}V_{\sigma,\sigma'}\rho_\sigma
\rho_{\sigma'}\right] \, ,\\
\ea
\hfill
\begin{picture}(3.375,0)
  \put(0,0){\line(1,0){3.375}}
  \put(0,0){\line(0,-1){0.08}}
\end{picture}

\begin{multicols}{2}
\noindent
where we have assumed short range interactions 
$V(\ell,\ell')=\delta(\ell-\ell')$. Note that the gauge potentials
appearing in the transformed Lagrangian are the local ${\cal A}_\mu$ given by 
Eq. (\ref{Aeff}).

We are interested in obtaining a low-energy effective description of the 
system under the assumption that our stiff string fluctuates over typical 
length and time scales which are long compared to the inverse of the Fermi 
wave-vector and Fermi energy of the one-dimensional 
electron gas.. In order to separate the fast and slow degrees of freedom 
we decompose the fermionic fields into left ($\eta=-1$) and right ($\eta=1$) 
moving fields, $\psi_{\eta,\sigma}$, which describe the slow 
long-wavelength variations around the respective Fermi points
\be
\label{psilr}
\psi_{\sigma}(x,t)=e^{-i\bar{k}_F x}\psi_{-1,\sigma}(x,t)
+e^{i\bar{k}_F x}\psi_{1,\sigma}(x,t) \; .
\ee
Here we have also used the assumption that the number of electrons 
on the string, $N_e$, is fixed and hence the average projected 
electronic density, $\bar{n}=N_e/L_x$, and with it the projected Fermi 
wave-vector, $\bar{k}_F=\pi \bar{n}/2$, are time independent and equal 
their values in the straight static string\cite{comment1}. 

The low-energy theory is obtained by using the decomposition, 
Eq. (\ref{psilr}), keeping leading order terms in $\bar{k}_F$,
which is assumed to be large, and neglecting irrelevant terms in the 
renormalization group sense. We also assume that the $2\bar{k}_F$ 
components of the gauge potentials and the metric vanish thus allowing
us to disregard single-particle backscattering. 

The theory can then be bosonized in the standard way\cite{emery1d,vondelft} 
with attention to the metric factors appearing in the kinetic term of 
Eq. (\ref{Ltrans}). The fermionic and bosonic representations are 
related via the identity
\be
\label{bidentity}
\psi_{\eta,\sigma}(x,t)=\frac{1}{\sqrt{2\pi a_e}}F_{\eta,\sigma}\exp[
-i\Phi_{\eta,\sigma}(x,t)] \; ,
\ee
where the self-dual fields $\Phi_{\eta,\sigma}$ are themselves 
combinations of the bosonic fields $\phi_c$ and $\phi_s$ 
and their conjugated momenta $\partial_x\theta_c$ and $\partial_x\theta_s$ 
\be
\label{decomp}
\Phi_{\eta,\sigma}=\sqrt{\frac{\pi}{2}}\,[(\theta_c-\eta\phi_c)
+\sigma(\theta_s-\eta\phi_s)] \; .
\ee
The Klein factors $F_{\eta,\sigma}$ in Eq. (\ref{bidentity}) are responsible 
for reproducing the correct anti-commutation relations between different 
fermionic species and $a_e\sim \bar{k}_F^{-1}$ is a short distance cutoff.

The result of these manipulations is the bosonic representation of the
effective electronic Lagrangian 

\end{multicols}
\widetext

\noindent
\setlength{\unitlength}{1in}
\begin{picture}(3.375,0)
  \put(0,0){\line(1,0){3.375}}
  \put(3.375,0){\line(0,1){0.08}}
\end{picture}
\ba
\label{bosL}
\nonumber
L_e=\int_0^{L_x} dx \Bigg\{&&
%\left(e{\cal A}_0-E_F-\frac{e^2}{2mc^2}
%{\cal A}_1^2\right)\sqrt{g}n +
\sum_{\alpha=c,s}\left[\partial_t\phi_\alpha\partial_x\theta_\alpha
-\frac{\widetilde{v}_\alpha \widetilde{K}_\alpha}{2}
(\partial_x\theta_\alpha)^2-\frac{\widetilde{v}_\alpha}
{2\widetilde{K}_\alpha}(\partial_x\phi_\alpha)^2\right] \\
%-\frac{1}{\sqrt{g}}\partial_t\ell\partial_x\theta_\alpha\partial_x\phi_\alpha
+&&\left[e{\cal A}_0-g^{-1}\bar{E}_F-
\frac{e^2}{2mc^2}g^{-1}{\cal A}_1^2-\frac{1}{32m}g^{-3}(\partial_x g)^2\right]
%-\frac{i}{4}\frac{\partial_t g}{g}+\frac{i}{4}\frac{\partial_x g} 
%{g^{3/2}}\partial_t\ell 
\left( \bar{n} + \sqrt{\frac{2}{\pi}} \,\partial_x\phi_c \right) 
+\bar{v}_F\frac{e}{c}\sqrt{\frac{2}{\pi}}g^{-1}{\cal A}_1
\partial_x\theta_c 
%-\frac{e^2}{2mc^2}{\cal A}_1^2 \sqrt{g}n
\Bigg\} \; ,
%+\left[v_F\frac{e}{c}{\cal A}_1 + m\ell\partial_t v_F+\frac{i}{4}
%v_F\frac{\partial_x g}{g^{3/2}}\right]\sqrt{\frac{2}{\pi}} 
%\,\partial_x\theta_c \Bigg\} \; .
\ea
\hfill
\begin{picture}(3.375,0)
  \put(0,0){\line(1,0){3.375}}
  \put(0,0){\line(0,-1){0.08}}
\end{picture}

\begin{multicols}{2}
\noindent
where $\bar{E}_F=\bar{k}_F^2/(2m)=m\bar{v}_F^2/2$ is
the Fermi energy of the non-interacting electron gas on the static system.
The ${\cal A}_1^2$ term was kept since its diamagnetic
contribution to the conductivity is essential if one is to obtain the 
correct response when the electric field is introduced via a
time-dependent vector potential, as we do later on. It also 
renormalizes the string dependent part of the scalar potential. 

The first two terms in Eq. (\ref{bosL}) involving the average projected 
electronic density, $\bar{n}$, are responsible for the renormalization 
of the bare string parameters. This becomes evident if we expand them 
to second order in derivatives of $Y$ in the absence of external 
electromagnetic fields. The first term ($e{\cal A}_0\bar{n}$) is 
identically $m\bar{n}\int_0^{L_x} dx (\partial_t Y)^2/2$. It 
expresses the fact that due to their constrained dynamics the electrons 
are dragged with the string as it moves, thereby increasing its effective 
mass density
\be
\label{renrho}
\rho\rightarrow\rho+m\bar{n} \; .
\ee
The second term reads $-\bar{E}_F\bar{n}+2\bar{E}_F
\bar{n}\int_0^{L_x} dx (\partial_x Y)^2/2$. It manifestly 
tells us that the electronic kinetic energy is lowered by the fluctuations. 
This gain in kinetic energy favors a more flexible string through the 
renormalization of its tension according to
\be
\label{rensigma}
\sigma\rightarrow\sigma-2\bar{E}_F\bar{n} \; .
\ee
In the following we assume that these effects are small and have already 
been taken into account when writing the string Lagrangian. In other words, 
that the parameters appearing in Eq. (\ref{L_s}) are already the 
renormalized ones.

The effects of string fluctuations are weaved into the rest of the 
the Lagrangian, Eq. (\ref{bosL}), as well. While it resembles that of 
a standard Luttinger model it is different in two important ways. 
First, the gauge potentials ${\cal A}_{\mu}$ are mixtures of the 
external electromagnetic potentials and the string dynamics according 
to Eq. (\ref{Aeff}). Secondly, the velocities 
$\widetilde{v}_\alpha$ and the Luttinger parameters 
$\widetilde{K}_\alpha$ are no longer 
mere constants but instead are functions of position and time.  
Allowing, in the usual way, for different interaction strengths
between left and right movers $V_{\sigma,\sigma'}=V_4\delta_{\eta,\eta'}
+V_2\delta_{\eta,-\eta'}-V_{1\parallel}
\delta_{\eta,-\eta'}\delta_{\sigma,\sigma'}$ but neglecting 
backscattering involving spin-flipping, we find
\ba
\label{tildev}
\nonumber
\widetilde{v}_c(x,t)&=&\frac{g^{-\frac{1}{2}}}{2\pi}
\sqrt{(2\pi \bar{v}_F g^{-\frac{1}{2}}+2V_4)^2-
(2V_2-V_{1\parallel})^2} \; , \\
\widetilde{v}_s(x,t)&=&\frac{g^{-\frac{1}{2}}}{2\pi}
\sqrt{(2\pi \bar{v}_F g^{-\frac{1}{2}})^2-
V_{1\parallel}^2} \; ,
\ea
and
\ba
\label{tildeK}
\nonumber
\widetilde{K}_c(x,t)&=&\sqrt{\frac{2\pi \bar{v}_F g^{-\frac{1}{2}}
+2V_4-2V_2+V_{1\parallel}}
{2\pi \bar{v}_F g^{-\frac{1}{2}}+2V_4+2V_2-V_{1\parallel}}} \; , \\
\widetilde{K}_s(x,t)&=&\sqrt{\frac{2\pi \bar{v}_F g^{-\frac{1}{2}}
+V_{1\parallel}}
{2\pi \bar{v}_F g^{-\frac{1}{2}}-V_{1\parallel}}} \; .
\ea
Note, however, that the model still exhibits spin-charge separation.

At this point we quantize the string. The effective low energy theory 
is then described by a path integral over the string and electronic 
configurations with respect to the combined action 
$S_s+S_e=\int dt (L_s+L_e)$, defined by Eqs. (\ref{L_s}) and (\ref{bosL}).

\section{Correlation Functions}
\label{correlations}

\subsection{Effective Action for Projected Correlation Functions}
\label{projected}

As long as one is interested in 
calculating correlation functions of electronic order parameters which 
can be expressed solely in terms of the $\psi_{\eta,\sigma}(x)$, such as 
the projected singlet pair annihilation operator $O_{SS}(x)=\sum_{\sigma}
\sigma\psi_{1,\sigma}(x)\psi_{-1,-\sigma}(x)$, one may integrate out the 
string from the problem. Our assumption 
regarding the smallness of the string fluctuations allows us to carry 
out this procedure perturbatively in the small dimensionless parameter, 
$\epsilon$, defined in Eq. (\ref{epsilon}). We begin by doing so in 
the absence of external electromagnetic fields. 

Using a cumulant expansion for the electronic action we obtain a
power series in derivatives of $Y(x,t)$, which can be readily averaged 
over the string dynamics, as described by the Lagrangian, Eq. (\ref{L_s}). 
%$\langle \frac{\partial Y}{\partial x_\mu} 
%(\tau,x)\frac{\partial Y}{\partial x_\nu}(0,0)\rangle=\frac{\epsilon}{2}
%i^{(1-\delta_{\mu\nu})}(1+\frac{\overline{x}_{\mu}-ix_\mu}{a})^{-2}
%+{\rm H.c.}$, which hold as long as $u/k_BT>x_{\mu},a>0$. 
%Here $x^0=iut=u\tau$, $x^1=x$ and $\overline{x}_{0,1}=x_{1,0}$.
To first order in $\epsilon$, and for periodic boundary 
conditions along the $x$ direction, the resulting 
effective electronic Lagrangian is of the ordinary Luttinger type. 
In imaginary time it reads 
\ba
\label{Leff}
\nonumber
L_e^{\rm eff}=\int_0^{L_x} dx &&
\sum_{\alpha=c,s}\bigg[-i\partial_\tau\phi_\alpha\partial_x\theta_\alpha  \\
&&+\frac{{v}_\alpha {K}_\alpha}{2}
(\partial_x\theta_\alpha)^2+\frac{{v}_\alpha}
{2{K}_\alpha}(\partial_x\phi_\alpha)^2\bigg] \; ,
\ea
where the velocities are renormalized away from their values 
$\bar{v}_\alpha$ in the straight static string of length $L_x$ according to 
\ba
\label{v}
%\nonumber
{v}_\alpha&=&\bar{v}_\alpha-\epsilon\left[\frac{\bar{v}_\alpha}{2}+
\frac{\bar{v}_F}{4}
\left(\frac{1}{\bar{K}_\alpha}+\bar{K}_\alpha\right)
\right] \; , 
%\bar{v}_s&=&v_s^0\left[1-\frac{\epsilon}{2}-\frac{\epsilon}{2}
%\left(\frac{v_F^0}{v_s^0}\right)^2\right] \; ,
\ea
and the Luttinger parameters acquire the values 
\ba
\label{k}
%\nonumber
{K}_\alpha&=&\bar{K}_\alpha-\frac{\epsilon}{4}
\frac{\bar{v}_F}{{\bar{v}_\alpha}}
\left(1-\bar{K}_\alpha^2\right) \; . 
%\bar{K}_s&=&K_s^0\left[1-\frac{\epsilon}{4}\frac{v_F^0}
%{v_s^0}\left(\frac{1}{K_s^0}-K_s^0\right)\right] \; .
\ea
Following our notation, the barred quantities characterize the system 
with no fluctuations and $\alpha=c,s$.

The Luttinger parameters constitute a measure of the relative size of 
the kinetic and potential energies of the one-dimensional electron 
gas. $K_c$ progressively deviates from 1 as the strength of the
interactions in the system is increased compared to its Fermi
velocity. The deviation is negative for repulsive interactions and 
positive in the presence of attraction between the electrons. $K_s$ 
behaves in a similar manner but with an opposite sign for the deviation.  
As seen from Eq. (\ref{k}) the string fluctuations have the effect
of increasing the relative size of the interactions in the system. 
The origin of this effect can be easily traced back to the different 
dependence of the kinetic and potential energies of the gas on the
length, $L$, of the string. While the kinetic energy scale as $L^{-2}$ 
the potential energy behaves as $L^{-1}$. Since the fluctuations 
increase the average length of string $\langle L\rangle=
\left(1+\frac{1}{2}\epsilon-\frac{3}{8}\epsilon^2+\cdots\right)L_x$
the above mentioned changes in $K_c$ and $K_s$ follow.

It is interesting to understand the effects of fluctuations beyond the 
simple elongation of the string which leads to Eqs. (\ref{v}) 
and (\ref{k}). To this end we express ${v}_\alpha$ 
and ${K}_\alpha$ in terms of the corresponding quantities 
$\overline{v}_\alpha$ and $\overline{K}_\alpha$ 
of a straight static string of length $\langle L\rangle$ - the average 
length of the fluctuating string, and with the same number of particles
$N_e$. This is also useful if one wishes to study the case where the average
density along the string is held fixed rather than the projected
density. To lowest order in $\epsilon$ we find  
\ba
\label{vbar}
%\nonumber
{v}_\alpha&=&\overline{v}_\alpha
\left(1-\frac{\epsilon}{2}\right) \; ,
\ea
where the proportionality factor is still geometrical in nature and 
stems from the fact that ${v}_\alpha$ expresses the
$x$-component of the velocities on the fluctuating string.

As expected the first order difference between ${K}_\alpha$ 
and $\overline{K}_\alpha$ vanishes and one needs to obtain the
effective action to second order in $\epsilon$. Ignoring non-local 
terms that are generated in the process and which are 
irrelevant in the renormalization group sense we find that 
the effective action retains the same form as Eq. (\ref{Leff}) with  
\ba
\label{kbar}
\nonumber
{K}_\alpha=\overline{K}_\alpha&+&\epsilon^2\Bigg{\{}\frac{1}{2}
\frac{\overline{v}_F}{\overline{v}_\alpha}\left(1-K_{\alpha}^2\right)
\\
\nonumber
&+&\frac{1}{2}\frac{u}{\overline{v}_\alpha}\left(\overline{k}_F a\right)^2
\!\left[4+\left(\frac{u}{\overline{v}_F}-\frac{\overline{v}_F}{u}\right)^2
\!\overline{K}_{\alpha}^2\right]\delta_{\alpha,c}\!\Bigg{\}} \; . \\
\ea 
The first term in the curly brackets is similar to the first order
term in Eq. (\ref{k}) with the exception that it has the effect of 
decreasing the relative strength of the interactions. It too can be
removed by an appropriate choice of the length of the straight static 
reference system. More importantly, the charge Luttinger parameter
acquires an additional contribution which tends to 
increase its value. It originates from the induced attractive
interaction due to the exchange of string waves between particles.

Since the effective action is that of an ordinary interacting
one-dimensional electron liquid the projected correlation functions
take, at zero temperature, the familiar power-law form 
\cite{emery1d}, with exponents which are determined by the effective
parameters, Eqs. (\ref{k},\ref{kbar}). 
In some cases, however, one is interested in calculating 
the correlation functions as functions of both coordinates in the  
embedding plane, not least since these are the quantities measured 
in experiments. We therefore consider this issue next.

\subsection{The Single Hole Green Function and the Tunneling Density of 
States}
\label{shgf}

The single hole Green function in the two-dimensional plane is defined 
as 

\be
\label{green}
G^<(\vec{r}_1,\vec{r}_2,t)=
\langle\Psi_\sigma^{\dagger}(\vec{r}_1,t)\Psi_\sigma(\vec{r}_2,0)\rangle \; .
\ee
The operator $\Psi_\sigma(\vec{r})$ annihilates  
an electron with spin $\sigma$ at point $\vec{r}$ in a plane of dimensions
$L_x\times L_y$ and the average is with respect to the string and electronic 
actions, $S_s+S_e=\int dt (L_s+L_e)$, Eqs. (\ref{L_s}) and (\ref{bosL}).

In the following we will consider the case where the string is fixed at 
a point such that its rigid translations are eliminated. In order to 
restore translational invariance of the Green function along the 
$y$-direction we average it over $\bar{y}=(y_1+y_2)/2$. The result 
\be
\label{avegreen}
\bar{G}^<(x,y,t)=\frac{1}{L_y}\int d\bar{y} \, G^<(\vec{r}_1,\vec{r}_2,t) 
\ee
is a function of the relative coordinates $x=x_1-x_2$ and $y=y_1-y_2$ 
only. 

Since the electrons are confined to the string, it must pass through the 
space-time point where the electron is to be created or annihilated in 
order for such a process to be possible. Therefore  
\be
\label{Psi}
\Psi_\sigma(\vec{r},t)=\psi_\sigma(x,t)\sqrt{a}\, 
\delta\left[y-Y(x,t)\right] \; . 
\ee
%where $\psi_\sigma(x,t)=[g(x,t)]^{1/4}\tilde\psi_{\sigma}[\ell(x,t),t]$.

Applying the decomposition, Eq. (\ref{psilr}), and expressing the 
$\delta$-function constraints through their integral representation 
the Green function of a right moving hole (the spin polarization is 
irrelevant here) can be written as  
\ba
\label{decompgreen}
\nonumber
\bar{G}^<(x,y,t)&=&\frac{a}{L_y}e^{-i\bar{k}_F x}
\int d\bar{y}\int_{-\infty}^{\infty} \frac{d\lambda_1}
{2\pi} \frac{d\lambda_2}{2\pi} \,  \\
\nonumber
&&\hspace{-1.2cm}\times \langle e^{i\lambda_1[Y(x,t)-y_1]} 
\psi_{11}^{\dagger}(x,t) 
e^{i\lambda_2[Y(0,0)-y_2]}\psi_{11}(0,0)\rangle \\
\nonumber
&=&\frac{a}{L_y}e^{-i\bar{k}_F x}
\int_{-\infty}^{\infty}\frac{d\lambda}{2\pi} \, 
 \\ 
&&\hspace{-1.2cm}\times \langle e^{-i\lambda y} e^{i\lambda Y(x,t)}
\psi_{11}^{\dagger}(x,t)e^{-i\lambda Y(0,0)}\psi_{11}(0,0)\rangle \; . 
\ea
In the last step we transformed to new integration variables 
$\lambda=(\lambda_1-\lambda_2)/2$ and $\Lambda=\lambda_1+\lambda_2$ 
and carried out the $\Lambda$ and $\bar{y}$ integrals.

%where we have used the fact that in the limit 
%$L_y\rightarrow\infty$ the $\bar{y}$ integration gives 
%$2\pi\delta(\lambda_1+\lambda_2)$.

We will calculate the Green function $\bar{G}^<$ for finite temperatures 
from the imaginary-time-ordered correlation function  
\ba
\label{finiteT}
\nonumber
\bar{G}&&^<(x,y,\tau)=
\frac{a}{L_y} e^{-i\bar{k}_F x}\frac{1}{Z}\int_{-\infty}^{\infty} 
\frac{d\lambda}{2\pi} \int {\cal D}Y{\cal D}\psi \,  \\
&&\times  e^{-i\lambda y}\psi^{*}(x,\tau)\psi(0,0)
e^{i\lambda[Y(x,\tau)-Y(0,0)]} e^{-[S_s+S_e]} \; , 
\ea
where $Z$ is the partition function. Concentrating on the $Y$ functional 
integration we have 
\ba
\label{Yint}
\nonumber
&&\frac{1}{Z}\int{\cal D}Y 
e^{i\lambda[Y(x,\tau)-Y(0,0)]}e^{-[S_s+S_e]} \\
\nonumber
&&=\frac{1}{Z}\int{\cal D}Y e^{i\sum_{q,\nu}{\cal J}^*(q,\nu)Y(q,\nu)} \\
\nonumber
&&\hspace{1.48cm}\times e^{-\frac{1}{2}\beta L_x \sigma\sum_{q,\nu}
\left(q^2+\frac{\nu^2}{u^2}\right)|Y(q,\nu)|^2}e^{-S_e(Y,\psi)}  \\
&&=e^{-\frac{1}{2\beta L_x \sigma}\sum_{q,\nu}
\left(q^2+\frac{\nu^2}{u^2}\right)^{-1}|{\cal J}(q,\nu)|^2}\frac{1}{Z'}
e^{-\int_0^\beta d\tau L_e^{\rm eff}} \;  , 
\ea
with ${\cal J}(q,\nu)=\lambda[e^{-i(qx-\nu\tau)}-1]$, and where the sums run 
over the Matsubara frequencies $\nu_n=\frac{2\pi}{\beta}n=2\pi T n$, 
($k_B=1$), and the allowed wave-vectors $q_n=\frac{2\pi}{L_x}n\neq 0$. 
In the last line of Eq. (\ref{Yint}) we have shifted $Y(q,\nu)\rightarrow
Y(q,\nu)+\frac{i}{\beta L_x\sigma}\left(q^2+\frac{\nu^2}{u^2}\right)
{\cal J}(q,\nu)$, and calculated the effective fermionic action to 
first order in $\epsilon$. The result, to this order, 
is identical to the effective action, Eq. (\ref{Leff}), 
and $Z'=\int{\cal D}\psi e^{-\int_0^\beta d\tau L_e^{\rm eff}}$. 

Carrying out the remaining functional integral over $\psi$ is a simple 
matter that gives $G_{1d}(x,\tau)$ - the finite temperature single hole 
Green function of a one-dimensional electron gas with velocities 
${v}_\alpha$, Eq. (\ref{v}), and Luttinger parameters 
${K}_\alpha$, Eq. (\ref{k}). Finally, performing the 
Gaussian integral over $\lambda$, taking $\tau>0$, and analytically 
continuing to the real time axis we obtain in the limit $L_x\rightarrow\infty$
\be
\label{greenres}
\bar{G}^<(x,y,t)=F(x,y,t)G_{1d}^<(x,t) \; ,
\ee
where
%\ba
%\label{F}
%\nonumber
%F(x,y&&,t)=\frac{1}{\sqrt{2\pi\epsilon a^2}} \\
%\nonumber
%&&\times\frac{e^{-y^2\left\{2\epsilon a^2
%\ln\left[\left(\frac{\lambda_{T,u}}{a}\right)^2 h_{-1}\left(\frac{x-ut}
%{\lambda_{T,u}}\right)h_{-1}^*\left(\frac{x+ut}{\lambda_{T,u}}\right)
%\right]\right\}^{-1}}}
%{\sqrt{\ln\left[\left(\frac{\lambda_{T,u}}{a}\right)^2 h_{-1}\left(
%\frac{x-ut}{\lambda_{T,u}}\right)h_{-1}^*\left(\frac{x+ut}{\lambda_{T,u}}
%\right)\right]}}  \; , \\
%\ea
\be
\label{F}
F(x,y,t)=\frac{a}{L_y}\frac{\exp\left[-y^2/ f(x,t)\right]}
{\sqrt{\pi f(x,t)}} \; ,
\ee
\ba
\label{f}
\nonumber
f(x,t)=2\epsilon a^2 &{\Bigg \{}&
\ln\left[-i\left(\frac{\lambda_{T,u}}{a}\right)
\sinh\!\left(\frac{x-ut+ia}{\lambda_{T,u}}\right)\right] \\
\nonumber
&+&\ln\left[i\left(\frac{\lambda_{T,u}}{a}\right)
\sinh\!\left(\frac{x+ut-ia}{\lambda_{T,u}}\right)\right]{\Bigg \}} \; , \\
%\nonumber
%&&\times\ln\!\left[\!\left(\frac{\lambda_{T,u}}{a}\right)^2 
%\!\sinh\!\left(\frac{x-ut+ia}{\lambda_{T,u}}\right)\sinh\!
%\left(\frac{x+ut-ia}{\lambda_{T,u}}\right)\!\right] , \\
\ea
and where the one-dimensional Green function is given by 
\ba
\label{G}
\nonumber
G_{1d}^<(x,t)=\frac{1}{2\pi a_e}&&e^{-i\bar{k}_F x}\left(\frac
{a_e}{\lambda_{T,c}}\right)^{2\gamma_c+\frac{1}{2}}\left(\frac{a_e}
{\lambda_{T,s}}\right)^{2\gamma_s+\frac{1}{2}} \\
\nonumber
\times \prod_{\alpha=c,s} &&\left[-i\sinh
\left(\frac{x-v_\alpha t+ia_e}{\lambda_{T,\alpha}}\right)\right]^
{-\gamma_\alpha-\frac{1}{2}} \\
&&\!\times \left[i\sinh\left(\frac{x+v_\alpha t-ia_e}
{\lambda_{T,\alpha}}\right)\right]^{-\gamma_\alpha} \; . 
\ea
%Here we have introduced the function
%\be
%\label{h}
%h_\gamma(x)=\left[-i\sinh(x+ia)\right]^{-\gamma} \; ,
%\ee
In the above we have introduced the thermal lengths
\be
\label{lambda}
\lambda_{T,u}=\frac{u}{\pi T} \;\;\; , \;\;\;
\lambda_{T,c,s}=\frac{v_{c,s}}{\pi T} \; ,
\ee
and the exponents $\gamma_{c,s}$, which are defined as
\be
\label{gamma}
\gamma_{c,s}=\frac{1}{8}({K}_{c,s}+{K}_{c,s}^{-1}-2) \; .
\ee
Note that to first order in $\epsilon$ these 
exponents are larger by $\frac{\epsilon}{32}\frac{\bar{v}_F}
{\bar{v}_{c,s}}(\bar{K}_{c,s}-\bar{K}_{c,s}^{-1})^2$
than their values on the static string. However, they still obey 
$\gamma_{c,s}=0$ for non-interacting electrons.

The effects of the string fluctuations, as given by the factor $F(x,y,t)$, 
introduce a Gaussian decay in $y$ to the Green function and reduce its 
amplitude at long times or large $x$ separations. In particular it means 
that in the presence of fluctuations the low energy $(T<\omega\ll u/a)$ 
tunneling density of states 
\be
\label{dos}
\bar\rho^<(\omega)=\int\frac{dt}{2\pi}e^{i\omega t}\bar{G}^<(0,0,t)\propto
\frac{\omega^{2(\gamma_c+\gamma_s)}}{\sqrt{|\ln(a \omega/u)|}}
\Theta(\omega) \; ,
\ee
is logarithmically suppressed compared to its behavior in the static Luttinger 
liquid, on top of the renormalization of its power-law exponent. 
Here and in the following the energy is measured relative to 
the Fermi energy on the straight string $\bar{E}_F$ \cite{shiftcom}.

\subsection{The Spectral Function}
\label{spectralfun}

Angle resolved photoemission spectroscopy (ARPES) measures the single
hole spectral function 
\ba
\label{shspectral}
\nonumber
A^<(\vec{k},\omega)=\frac{1}{L_xL_y}\int\! d^2r d^2r'&& dt \,
e^{i[(\vec{k}\cdot(\vec{r}-\vec{r}')-\omega t]} G^<(\vec{r},\vec{r}\,' ,t)
\, . \\
\ea
%The Integral over the center of mass $y$-coordinate gives 
%$\bar{G}^<(x-x',y-y',t)$ and the integral over the relative 
%$y$-coordinate is simple owing to the Gaussian form of $F$, 
%Eq. (\ref{F}). The result of performing the $y$ Fourier transfom, 
%which we denote by $A^<(x,t;k_y)$, is then
The integrals over the $y$ coordinates are immediate owing to the 
$\delta$-function factors in $G^<(\vec{r},\vec{r}\,' ,t)$. One therefore 
finds 
\ba
\label{decompspec}
\nonumber
A^<(\vec{k},\omega)=\frac{a}{L_y}
&&\int dx dt\, e^{i[(k_x-\bar{k}_F)x-\omega t]} \\ 
\nonumber
&&\times \langle  e^{ik_yY(x,t)}
\psi_{11}^{\dagger}(x,t)e^{-ik_yY(0,0)}\psi_{11}(0,0)\rangle \; . \\
\ea
The average that appears in Eq. (\ref{decompspec}), which we denote by 
$A^<(x,t;k_y)$, can be evaluated for finite temperatures using similar 
manipulations as the ones indicated following Eqs. (\ref{finiteT}) and 
(\ref{Yint}). The result 
\be
\label{A(x,t)}
A^<(x,t;k_y)=F(x,t;k_y)G_{1d}^<(x,t) \; ,
\ee
where 
\ba
\label{Fk}
\nonumber
F(x,t;k_y)&=&\frac{a}{L_y}\left(\frac{a}{\lambda_{T,u}}\right)^
{2\Delta(k_y)} \\
\nonumber
&\times&\left[-i\sinh\left(\frac{x-ut+ia}{\lambda_{T,u}}\right)\right]
^{-\Delta(k_y)} \\
&\times& \left[i \sinh\left(\frac{x+ut-ia}{\lambda_{T,u}}\right)\right]
^{-\Delta(k_y)} \; , 
\ea
depends on $k_y$ through the exponent
\be
\label{delta}
\Delta(k_y)=\frac{\epsilon}{2}(a k_y)^2 \; .
\ee

To find the spectral function $A^<(\vec{k},\omega)$ we still need to 
perform the Fourier transform over $A^<(x,t;k_y)$, which can be written as 
a convolution of the Fourier transform of $F(x,t;k_y)$ and the one-dimensional 
spectral function $A_{1d}^<(k_x,\omega)$ 
\be
\label{convol}
A^<(\vec{k},\omega)=\frac{1}{(2\pi)^2}\int dq d\nu F(q,\nu;k_y)
A_{1d}^<(k_x-q,\omega-\nu) \; .
\ee
The finite-temperature one-dimensional spectral function has been evaluated 
in Ref. \ref{spectralref}. Using a similar technique and in terms of the 
dimensionless scaling variables
\be 
\label{scalevar}
\tilde{k}_x=\frac{uk_x}{\pi T} \;\;\; , \;\;\; 
\tilde{\omega}=\frac{\omega}{\pi T} \; ,
\ee
and the Fourier transform of $\lim_{a\rightarrow 0}
[-i\sinh(x+ia)]^{-\gamma}$ 
\be
\label{h(k)}
h_\gamma(k)={\rm Re}\left[(2i)^\gamma{\rm B}\left(\frac{\gamma-ik}{2},1-\gamma
\right)\right] \; ,
\ee
where ${\rm B}(x,y)$ is the beta function, one obtains 
\ba
\label{F(k,w)}
\nonumber
F(k_x,\omega;k_y)&=&\frac{a^3}{2uL_y}\left(\frac{a}{\lambda_{T,u}}\right)^
{2\Delta(k_y)-2} \\ 
&\times&\! h_{\Delta(k_y)}\!\left(\frac{\tilde{\omega}+\tilde{k}_x}{2}\right)
h_{\Delta(k_y)}\!\left(\frac{\tilde{\omega}-\tilde{k}_x}{2}\right)  . \!
\ea 

Evaluating the convolution is difficult in the general case, especially 
as $A_{1d}^<$ itself is a convolution of charge and spin pieces 
\cite{spectral}. However, some features of the spectral function 
may be deduced from general kinematical considerations without 
resorting to detailed calculations\cite{frac,ericaproc}.

\begin{figure}[ttt]
\narrowtext
\begin{center}
\leavevmode
%\vspace{.2cm}
\noindent
\hspace{0.3 in}
\centerline{\epsfxsize=2.7in \epsffile{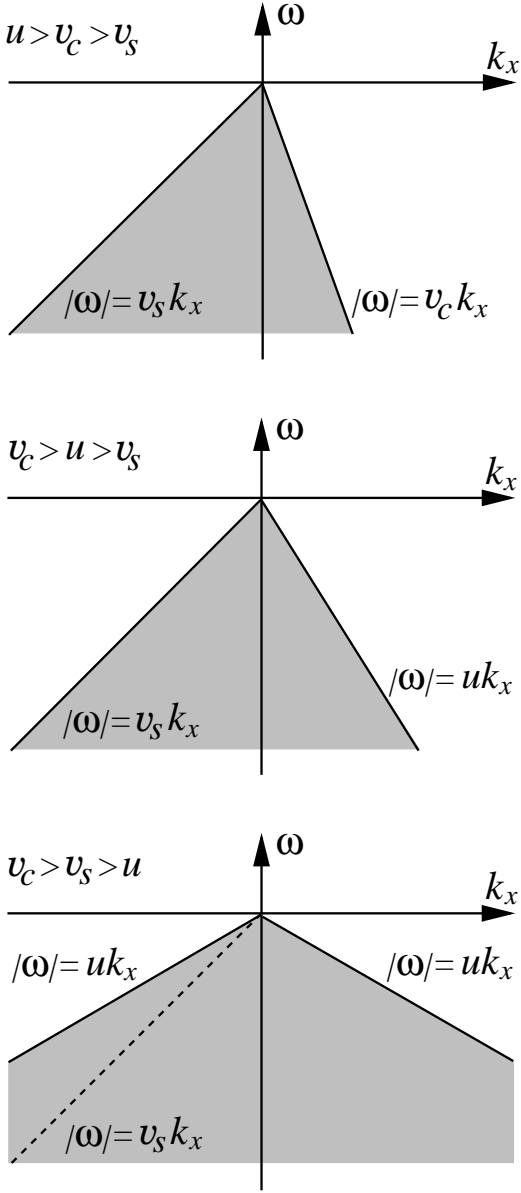}}
%\epsfxsize=3.3in
%\epsfysize=2.in
%\epsfbox{}
%\vspace{.2cm}
\end{center}
\caption
{The support of the spectral function, $A^<(\vec{k},\omega)$, of a right 
moving hole for various magnitudes of the string sound velocity $u$. 
At zero temperature $A^<(\vec{k},\omega)$ is nonzero only in the
shaded region of the $(k_x,\omega)$ plane. This figure assumes $k_y\neq 0$, 
a spin-rotationally invariant system and $v_c>v_s$. For moderate 
interactions and small string fluctuations most of the spectral 
weight is concentrated in the region next to the line $|\omega|=v_s k_x$.}
\label{fig1}
\end{figure}

An example is the the shape of the zero temperature support 
of $A^<(\vec{k},\omega)$, \ie the region in $(\vec{k},\omega)$ space where 
it is non-vanishing at $T=0$. 
Because of the spin-charge separation that takes place in one dimension 
the addition of an electron to the system necessarily involves the creation 
of at least one charge excitation and one spin excitation. When the string 
itself is a dynamic entity this process also involves the emission of 
string waves. The momentum $k_x$ and energy $\omega$ of the added electron 
are distributed among these excitations

\be
\label{kinematics}
k_x=k_c+k_s+k_u \;\;\; , \;\;\; \omega=v_c|k_c|+v_s|k_s|+u|k_u| \; .
\ee
It is easy to check that any point in the $(k_x,\omega)$ plane above
the dispersion curve of the slowest excitation branch may be reached
by appropriately distributing the momentum and energy of the added 
electron in a manner consistent with the conservation laws, Eq.
(\ref{kinematics}). If the system possesses additional symmetries, 
such as spin-rotation symmetry that inhibits the decay of a right 
moving electron into left moving spin excitations, \ie enforces $k_s>0$, 
the region of support may be further constrained. The spectral function 
of a hole is subject to a similar consideration with results which 
are presented in Fig. \ref{fig1}. Note that in this figure and in the 
following $k_x$ and $\omega$ are measured relative to $\bar{k}_F$
and $\bar{E}_F$ respectively. 

While the support is governed by kinematics the distribution of the 
spectral weight inside this region is determined by the matrix elements 
which connect the single hole state with various multi-excitation states 
which form the continuum. For weakly interacting electrons much of the 
weight is concentrated around the non-interacting dispersion line 
$|\omega|=v_sk_x \approx v_c k_x$. It progressively spreads 
throughout the region of support with increasing strength of the 
electron-electron interactions. It is also smeared due to the 
excitation of string waves. This smearing is more pronounced for 
larger values of $k_y$. It is a result of the fact that the 
$y$ dependent parts of the electronic operators in the spectral 
function, Eq. (\ref{decompspec}), shift the momentum of the string 
by $k_y$, thus increasing the number of string waves accordingly.

Analytical progress in evaluating the spectral function is possible for 
several special cases:

\subsubsection{The case $u=v_c=v_s$} 
Denoting $\lambda_T=\lambda_{T,u}$ and $\gamma=\gamma_c+\gamma_s$ one finds 
\ba
\label{equalv}
\nonumber
A^<(\vec{k},\omega)&=&\frac{1}{4\pi^2}\frac{1}{T}\frac{a}{L_y}
\left(\frac{a_e}{\lambda_T}
\right)^{2\gamma}\left(\frac{a}{\lambda_T}\right)^{2\Delta(k_y)} \\ 
\nonumber
&\times&h_{\gamma+\Delta(k_y)+1}\!\left(\frac{\tilde{\omega}+\tilde{k}_x}{2}
\right)h_{\gamma+\Delta(k_y)}\!\left(\frac{\tilde{\omega}-\tilde{k}_x}{2}
\right) . \\
\ea
The zero temperature limit of this result is easily obtained using
\be
\label{ztemp}
h_\gamma(|k|\rightarrow\infty)=\frac{2\pi}{\Gamma(\gamma)}\Theta(-k)
(-k)^{\gamma-1} \; .
\ee
 
\subsubsection{The case $u=v_c\neq v_s$ and $\gamma_s=0$} 
This case is of importance since $\gamma_s=0$ for a spin-rotationally 
invariant system. Denoting by $r=v_s/v_c$ the ratio between the spin 
and charge velocities one finds
\ba
\label{spinrotinv}
\nonumber
A^<(\vec{k}&&,\omega) = \frac{1}{(2\pi)^3\sqrt{r}T}\frac{a}{L_y}
\left(\frac{a_e}{\lambda_{T_c}}\right)^{2\gamma_c}
\left(\frac{a}{\lambda_{T_c}}\right)^{2\Delta(k_y)} \\
\nonumber
&&\times\int_{-\infty}^{\infty} dq 
\, h_{\frac{1}{2}}(q) \, h_{\gamma_c+\Delta(k_y)}\!
\left[\frac{\tilde{\omega}-\tilde{k}_x}{2}
-\left(1-\frac{1}{r}\right)\frac{q}{2}\right] \\
\nonumber
&&\hspace{1.75cm}
\times h_{\gamma_c+\Delta(k_y)+\frac{1}{2}}\!\left[\frac{\tilde{\omega}
+\tilde{k}_x}{2}-\left(1+\frac{1}{r}\right)\frac{q}{2}\right]  \; .\\
%A^<(\vec{k}&&,\omega) = \frac{\sqrt{r}}{4\pi^3}\frac{1}{T}\frac{a}{L_y}
%\left(\frac{a_e}{\lambda_{T_c}}\right)^{2\gamma_c}
%\left(\frac{a}{\lambda_{T_c}}\right)^{2\Delta(k_y)} \\
%\nonumber
%&&\times\int_{-\infty}^{\infty} dq 
%\,h_{\gamma_c+\Delta(k_y)+\frac{1}{2}}\left[\frac{\tilde{\omega}
%-r\tilde{k}_x}{2}+(1+r)q\right] \\
%\nonumber
%&&\times h_{\gamma_c+\Delta(k_y)}
%\left[\frac{\tilde{\omega}-r\tilde{k}_x}{2}
%-(1-r)q\right]h_{\frac{1}{2}}\left[r\tilde{k}_x-2rq\right]  . \\
\ea

The results, Eqs. (\ref{equalv}) and (\ref{spinrotinv}), are similar
to the spectral functions of a static one-dimensional electron gas 
under the same conditions \cite{spectral} with the exception that the 
exponents that govern their behavior are now functions of the strength 
of the string fluctuations and, more importantly, of $k_y$, as given by 
the definition of  $\Delta(k_y)$, Eq. (\ref{delta}). 
We expect this similarity to continue to exist as long as $u$ is 
not very different from either $v_c$ or $v_s$.

If one relaxes the constraint that confines the electronic 
wave function to live strictly on the one-dimensional string and allows 
some leakage of it into the surrounding environment the above spectral 
functions are modified. In the simple case where one can replace the 
electronic operator, Eq. (\ref{Psi}), by $\Psi_\sigma(\vec{r},t)=
\psi_\sigma(x,t)\varphi[y-Y(x,t)]$, they turn, in real space, into 
a convolution of the results obtained with the $\delta-$function
constraint and the wave function in the transverse direction 
$\varphi(y)$. Consequently, in Fourier space the spectral functions 
are multiplied by $|a^{-1/2}\int dy e^{-ik_y y}\varphi(y)|^2$.

\subsection{Optical Conductivity}

The effective electronic Lagrangian, Eq. (\ref{bosL}), also contains the 
electromagnetic response of the system. In particular, we are interested in 
its optical conductivity, {\it i.e.}, the manner in which it responds to the 
application of a time-dependent spatially-uniform external electric field.   

%To this end we consider a system of $N_s=L_y/\lambda$ free (in this section 
%we allow rigid translations), independent
%strings positioned at points $y_i$ along the $y$-axis which are separated 
%by a distance $\lambda$ from each other \cite{transcom}. The action
%now becomes a sum of $N_s$ terms of the form (\ref{bosL}), where in 
%the term describing the i-th string the external potentials are to be 
%evaluated at the space-time coordinates $[x,y_i+Y_i(x,t),t]$. In the 
%continuum limit $\sum_{i=1}^{N_s}\rightarrow \lambda^{-1}\int_0^{L_y} dy$,
%and the fields become functions of the continuous coordinate $y$: 
%$Y_i(x,t)\rightarrow Y(x,y,t)$ etc.

To this end we consider a free string (in this section 
we allow the string to perform rigid translations) and introduce the 
electric field via a time dependent vector potential $\vec{A}$ which enters 
the electronic Lagrangian, Eq. (\ref{bosL}), through the effective 
vector potential ${\cal A}_1$, Eq. (\ref{Aeff}).
%are to be evaluated at the space-time coordinates $[x,Y(x,t),t]$.
We evaluate the (imaginary time) generating functional 
$e^{-S[\vec{A}(i\omega)]}$ by integrating out the string
and electronic degrees of freedom using a cumulant expansion to first 
order in $\epsilon$ and second order in $\vec{A}$. The optical
conductivity is then given by 
\be
\label{optical}
\sigma_{\mu\nu}(i\omega)=\frac{c^2}{\beta L_x L_y}\frac{1}{\omega}
\frac{d^2 S[\vec{A}(i\omega)]}{dA_\mu(-i\omega)dA_\nu(i\omega)} \; .
\ee
By analytically continuing $i\omega\rightarrow \omega+i\delta$ we find 
in the limit of zero temperature and $a\rightarrow 0$ (while keeping 
$\epsilon$ and $\bar{n}a$ finite)
\ba
\label{sigmaxx}
\sigma_{xx}(\omega)&=&\frac{2}{\pi}\frac{e^2}{L_y}\bar{v}_F
(1-\epsilon)\frac{i}{\omega+i\delta} \; , \\
\nonumber
\sigma_{yy}(\omega)&=&2\pi\frac{e^2}{L_y}(\bar{n}a)^2
u\epsilon\frac{i}{\omega+i\delta} \\
\label{sigmayy}
&&+\frac{2}{\pi}\frac{e^2}{L_y}\frac{u\bar{K}_c+\bar{v}_F}
{u+\bar{v}_c}\left(\bar{v}_c-\frac{\bar{v}_F}
{\bar{K}_c}\right)\epsilon \frac{i}{\omega+i\delta} \; .
\ea
In a system with momentum-independent interactions 
\cite{neutral} $\bar{v}_F=\bar{v}_c\bar{K}_c$ and as 
a result 
\ba
\label{misigmaxx}
\sigma_{xx}(\omega)&=&\frac{2}{\pi}\frac{e^2}{L_y}{v}_c
{K}_c\frac{i}{\omega+i\delta} \; , \\
\label{misigmayy}
\sigma_{yy}(\omega)&=&\frac{8}{\pi}\frac{e^2}{L_y}(\bar{k}_Fa)^2
u\epsilon\frac{i}{\omega+i\delta} \; .
\ea

The expression for $\sigma_{xx}$ is similar to the optical conductivity 
of a static Luttinger liquid \cite{schulz} but with renormalized
parameters owing to the effects of string fluctuations. 

The result for $\sigma_{yy}$ is the conductivity of a classical string 
charged with immobile charges of linear density $\bar{n}$. One can 
simply derive it by solving the classical equation of motion of 
a charged string in the presence of an electric field: 
$\rho d^2Y/dt^2 - \sigma d^2Y/dx^2 +\bar{n}eE_y(t)=0$   
and using the fact that the current density in the $y$-direction 
is given by $J_y(t)=(-e\bar{n}/L_y)(dY/dt)$.

The effects coming from the motion of the charges along the string as
it fluctuates vanish in the limit $a\rightarrow 0$ and $\epsilon$ fixed. 
However, in case the string is fixed at a point such that it can not 
execute rigid translations in the $y$-direction the result 
for $\sigma_{yy}$, Eq. (\ref{misigmayy}), vanishes while the contributions 
to $\sigma_{yy}$ coming from the motion of charge along the string 
remain and dominate. To lowest order in $\epsilon$ and $a$ we find
them to be (in the case of momentum-independent interactions)
\ba
\label{sigmayypinned}
\nonumber
&&\sigma_{yy}(\omega)=\frac{1}{\pi}\frac{e^2}{L_y}\frac{\bar{K}_c}
{u+\bar{v}_c}\epsilon a^2\left[ \pi|\omega|-2i\omega\ln\left(
\frac{u+\bar{v}_c}{e^{\gamma}a|\omega|}\right)\right]\;  , \\
\ea
where here $\gamma$ is the Euler constant. In this case the real part of  
$\sigma_{yy}$, originating from the excitation of string waves\cite{comcond}, 
is linear in $\omega$ with an oscillator strength (integrated up to the 
cut-off $u/a$) which is smaller by a factor $\epsilon(u/v_c)^2$ than the 
weight of the Drude-peak in $\sigma_{xx}$.

\subsection{A String in a Confining Potential} 

Next we consider the case in which the string is placed inside a parabolic 
confining potential. The string Lagrangian becomes 
\be
\label{Lscon}
L_s= \int_0^{L_x} dx \, {{\sigma}\over{2}}\left[\frac{1}{u^2}
\left({\partial Y}\over{\partial t}\right)^2 - 
\left({\partial Y}\over{\partial x}\right)^2 - \kappa^2 Y^2 \right ] \; .
\ee

Since the renormalization of the parameters of the effective action 
for calculating projected correlations is governed by the 
short-wavelength string fluctuations the results,
Eqs. (\ref{Leff})-(\ref{kbar}), are unchanged by the long-wavelength cutoff 
$\kappa^{-1}$ set by the confining potential (as long as $\kappa^{-1}\gg a$).  

The modified string propagator does affect the result for the single hole  
Green function [see Eq. (\ref{Yint}).] At zero temperature we find that the 
function $f(x,t)$ which enters the factor $F(x,y,t)$ in
Eq. (\ref{F}) is changed into
%\ba
%\label{modF}
%\nonumber
%F(x,&&y,t)=\frac{1}{2\sqrt{\pi\epsilon a^2}} \\
%\nonumber
%&&\times\frac{e^{y^2\left\{4\epsilon a^2
%\left[\ln(c \kappa a)+K_0[\kappa
%\sqrt{(x-ut+ia)(x-ut-ia)}]\right]\right\}^{-1}}}
%{\sqrt{\left|\ln(c\kappa a)+K_0[\kappa
%\sqrt{(x-ut+ia)(x-ut-ia)}]\right|}}  \; , \\
%\ea
\ba
\label{modf}
\nonumber
f(x&&,t)=4\epsilon a^2 \\ 
\nonumber
&&\times\left[K_0(\kappa a)
-K_0\!\left(\kappa\sqrt{(x-ut+ia)(x+ut-ia)}\right)\right] , \\
\ea
where $K_0$ is the modified Bessel function.
Consequently the low energy tunneling density of states exhibits a 
crossover from the Luttinger liquid behavior 
$\rho(\omega)\propto \omega^{2(\gamma_c+\gamma_s)}$ for $\omega<\kappa u$ 
to the logarithmically suppressed form, Eq. (\ref{dos}), at higher 
frequencies.

Similarly, after performing the $y$ Fourier transform, one obtains  
\ba
\label{modspec}
\nonumber
&&A^<(x,t;k_y)=\frac{a}{2\pi L_y}e^{-i\bar{k}_F x}
a_e^{2(\gamma_c+\gamma_s)} e^{-2\Delta(k_y)K_0(\kappa a)} \\
\nonumber
&& \;\; \times \;  \exp \left[2\Delta(k_y) K_0\!\left(\kappa
\sqrt{(x-ut+ia)(x+ut-ia)}\right)\right] \\
\nonumber
&& \;\; \times \!\prod_{\alpha=c,s} [-i(x-v_\alpha t+ia_e)]^{-(\gamma_\alpha
+\frac{1}{2})} [i(x+v_\alpha t-ia_e)]^{-\gamma_\alpha} , \\
\ea
where $\gamma_c$, $\gamma_s$ and $\Delta(k_y)$ retain their previously
found values, Eqs. (\ref{gamma},\ref{delta}). 
Although we are unable to provide a closed expression
for $A^<(\vec{k},\omega)$ we expect it to exhibit a crossover around 
$\omega\sim -u\sqrt{k_x^2+\kappa^2}$ where the exponent of its
approximate power-law behavior increases by $\Delta(k_y)$ relative to
its value in the region $\omega\approx-u|k_x|$. 

Finally, while the result for $\sigma_{xx}(\omega)$, Eq. (\ref{misigmaxx}),
is unchanged by the presence of the confining potential the optical 
conductivity along the $y$ direction becomes
\be
\label{consigmayy}
\sigma_{yy}(\omega)=\frac{8}{\pi}\frac{e^2}{L_y}(\bar{k}_F a)^2
u\epsilon\frac{i\omega}{(\omega+i\delta)^2-(u\kappa)^2} \; .
\ee
This is again just the classical conductivity of a charged string
fluctuating in a parabolic well and can be easily derived by solving 
the classical equation of motion in a similar fashion to the one we
have outlined following Eq. (\ref{misigmayy}).

\section{Discussion}
\label{discussion}

As we mentioned in the Introduction our model was motivated by the growing 
body of data which suggests that striped inhomogeneous states appear over 
a wide range of the phase diagram of the cuprate high temperature 
superconductors \cite{ourreview}. Our aim here is to explore possible 
connections between the model we have studied and a few aspects of the 
phenomenology of these systems. There is no doubt that the physics of 
stripes in doped Mott insulators is more involved than the one we have 
considered. The dynamics of stripes as loci of sites visited by holes 
is intricately and self-consistently determined by the dynamics of the holes, 
their interactions with the spin background and with each other. 
In our model we have concentrated on a single string and assumed that 
rotational invariance has been completely broken, possibly due to a strong 
orienting field originating from the underlying lattice. A more realistic 
treatment of smectic and nematic stripe phases in the cuprates should 
include the interactions between stripes, which are expected 
to be invariant under small rotations \cite{qhsmectic1}. 
We defer the study of these issues to the future. Notwithstanding, 
we hope that the model captures at least 
some of the features associated with the relevant stripe physics.   

\subsection{The Pseudogap}

One such feature, which is revealed by the model, is the tendency of the 
holes to enhance the flexibility of the string in order to increase its 
fluctuations and by that the gain in their kinetic energy. We believe that 
this is also the driving force behind stripe fluctuations in the coper-oxygen 
planes, where holes hop in the transverse direction to the stripe in order to 
lower their kinetic energy. As we have shown, as long as the number of
electrons is fixed, an increase in the string 
fluctuations leads to an increase in the relative strength of the 
electron-electron interactions as reflected by the Luttinger 
parameters $K_\alpha$, Eq. (\ref{k}). This fact may carry with it 
consequences to the size of the pseudogap which is observed, 
especially in underdoped samples\cite{timusk,loram}.

A possible explanation of the pseudogap phenomenon is the so-called 
``spin gap proximity effect'' suggested in Ref. \ref{spingapref}. In this 
scenario two coupled one-dimensional systems with different Fermi 
wave-vectors are considered, modeling the stripe and its environment. 
Single particle tunneling between the systems is suppressed owing to
the difference in their Fermi wave-vectors. However, under appropriate 
circumstances singlet pair-tunneling processes can become relevant. 
When this happens the coupled system scales to a new strong coupling 
fixed point which exhibits a total spin gap and strong global 
superconducting fluctuations. The physics is analogous to the
proximity effect in conventional superconductors since both are 
driven by the gain in the zero point kinetic energy of the pairs 
which outweighs the cost of pairing. 

The relevancy of the pair-tunneling term is governed by its scaling 
dimension
\ba
\nonumber
\delta_{pair}=\frac{1}{2}\left(\frac{A}{K_c}+\frac{B}{K_c^{(e)}}+K_s 
+K_s^{(e)} \right) \; , 
\ea
where $K_\alpha$, $K_\alpha^{(e)}$ are the Luttinger parameters of the 
stripe and the environment respectively. $A=1$ and $B=1$ in the absence 
of forward scattering inter-system density-density and current-current 
interactions but are in general complicated functions of the coupling 
constants. In particular they are decreasing functions of $K_c$. 
Pair tunneling is perturbatively relevant if $\delta_{pair}<2$ and 
irrelevant otherwise. 

Intra-stripe repulsive interactions ($K_c<1$) increase the value of 
$\delta_{pair}$, thus making it less relevant. This is physically 
reasonable since repulsive interactions within the stripe are
unfavorable for pairing and therefore for pair tunneling. 
Since stripe fluctuations have 
the effect of decreasing the value of $K_c$, see Eq. (\ref{k}),
they tend to increase the scaling dimension $\delta_{pair}$ and 
thus reduce the effectiveness of the spin-gap proximity effect. 
Since there is experimental evidence that correlates between higher levels 
of doping and a greater degree of stripe fluctuations this would mean 
that fluctuations may account, at least partially, for the decrease 
in the pseudogap with doping.

Here it should be noted, however, that fluctuations have additional 
effects on the Josephson coupling between stripes. As  
this coupling involves tunneling of pairs from one system to another 
it is governed by the points of closest approach between them. Since 
fluctuations increase the probability of having the two systems in 
close proximity the amplitude of such tunneling processes is exponentially 
enhanced by them\cite{kfe}. In addition it has been shown that 
inter-stripe forward scattering interactions tend to increase the relevance 
of pair tunneling\cite{sliding1,sliding2}. Such interactions arise naturally 
when integrating out the shape fluctuations in coupled chain systems. 

Which of the three effects just mentioned dominates the physics is a matter 
of details. It is possible that in the spin-gap proximity effect, in which 
pairs do not have to tunnel through an intervening barrier and where 
the forward scattering interactions induced by the fluctuations are 
second order in $\epsilon$,  
the reduction in the kinetic energy along the stripe is the most 
important, at least for small fluctuations. It is also likely that for 
the establishment of global phase coherence in a quasi-one-dimensional 
superconductor through Josephson tunneling between stripes the
increase of the tunneling amplitude due to fluctuations is the major effect.  

\subsection{Angle Resolved Photoemission Spectroscopy}

ARPES measurements provide indirect evidence for the existence of stripes 
in the cuprate compounds \cite{arpesreview}. For instance, 
measurements \cite{zhou01} of \LSCO and the related materials 
\cite{zhou01,zhou99}  \LNSCOsc and \LNSCOin have revealed 
that the frequency-integrated spectral weight is confined inside 
one-dimensional segments in momentum space. These segments can be 
interpreted as the Fermi surfaces of quarter-filled one-dimensional stripes  
running along the $a$ and $b$ directions in the planes. 

The spectral function $A^<(\vec{k},\omega)$, which is measured by ARPES, 
is often a broad function of the frequency $\omega$, especially in 
the vicinity of the anti-nodal points $(0,\pm\pi)$, $(\pm\pi,0)$. 
This behavior is naturally explained \cite{frac} if one considers 
the signal as coming from a collection of one-dimensional systems 
which are capable of producing such wide spectra for a large enough 
interaction strength [large enough values of $\gamma$, 
see Eq. (\ref{gamma})]. 

The width of any $\omega$ structure in $A^<(\vec{k},\omega)$ typically 
becomes smaller as one moves from the anti-nodal regions toward the 
$\Gamma$-point $(0,0)$ or the nodal regions around $(\pm\pi/2,\pm\pi/2)$.
This tendency is ubiquitous and appears in various systems, for example 
\cite{valla} \BSCCO. Within the interpretation of the two-dimensional 
Fermi surface as being a superposition of two Fermi
surfaces of one-dimensional stripes, the anti-nodal regions correspond 
to wave-vectors with a large component in the direction perpendicular
to the stripes. We have seen that the effects of stripes fluctuations 
tend to increase, via the term $\Delta(k_y)=\epsilon(ak_y)^2/2$, the 
value of the exponent that governs the behavior of the spectral
function, Eqs. (\ref{equalv},\ref{spinrotinv}). This enhancement grows 
with the magnitude of the transverse wave-vector component, $k_y$, 
and leads to broader spectra. It may constitute at least a partial 
reason for the observed broad spectra in the anti-nodal regions.

\subsection{Anisotropies in the Optical Conductivity}

We have shown that stripe fluctuations lead to an anisotropy in the optical 
conductivity, Eqs. (\ref{misigmaxx},\ref{misigmayy}), whose size is inversely 
proportional to the degree of fluctuations  as given by the parameter 
$\epsilon$. Such anisotropies of the far infra-red conductivity in the 
$a$-$b$ plane have been observed\cite{basov} in detwinned samples of 
\YBCO and \124. More recently similar anisotropies have been found
\cite{basovprivate} in the infra-red conductivity of a detwinned 
$x=0.03$ crystal of \LSCO. Although it is likely that part of the 
anisotropies observed in YBCO are due to the Cu-O chains present 
in this system it is nevertheless probable, particularly in view of 
the neutron scattering data, that at least some of the effect comes 
from stripes. If this is the case then the fact that the anisotropy is 
about a factor of 2-3 at most, teaches us that in the measured samples 
there is a large degree of stripe orientational disorder and fluctuations.

\acknowledgments
It is a pleasure to thank Oded Agam, Shmuel Elitzur, Eduardo Fradkin, 
Steve Kivelson and Moshe Schechter for 
helpful discussions. This research was supported by the Israel Science 
Foundation (grant No. 193/02-1).

\end{multicols}


\begin{references}
\bibitem{acomment} In this paper we employ exponential cut-off functions 
$e^{-|k|a}$ and $e^{-|\omega|a/u}$ 
in the momentum and frequency integrals over the string fluctuations.
\bibitem{comment1} Note that this also means that the 
decoherence effect, which was proposed in Ref. \ref{kferef} in order to 
reduce the inter-stripe charge density-wave coupling, is absent if the
number of electrons on each stripe is conserved, as being assumed here.
\bibitem{shiftcom} The shift in the energy originates from the time 
dependence of the Klein factor $F_\eta$ in $\psi_\eta$. In writing the
effective electronic Hamiltonian  we have dropped a term 
$\frac{m}{2}\bar{v}_F^2\hat{N}_\eta=\bar{E}_F\hat{N}_\eta$ 
which accounts for the energy needed to add a particle to the system. 
Since the normal ordered number operator $\hat{N}_\eta$ obeys \cite{vondelft}
$[N_\eta,F_\eta^\dagger]=F_\eta^\dagger$ this term induces a time dependence  
$F_\eta^\dagger (t)=e^{i\bar{E}_F t}F_\eta^\dagger$.
\end{references}
\end{document}